\documentclass{article}
\usepackage{spconf,amsmath,amssymb,graphicx, booktabs, multirow}

\newcommand{\share}[1]{\langle{#1}\rangle}

\title{Privacy-preserving Automatic Speaker Diarization}

\name{Francisco Teixeira$^1$, Alberto Abad$^1$, Bhiksha Raj$^{2,3}$, Isabel Trancoso$^1$\thanks{This work was supported by Portuguese national funds through Fundação para a Ciência e a Tecnologia (FCT), with references UIDB/50021/2020 and CMU/TIC/0069/2019, as well as by the Portuguese Recovery and Resilience Plan (RRP) through project C645008882-00000055 (Responsible.AI).
}}
\address{$^1$INESC-ID/IST, University of Lisbon, Portugal, $^2$LTI, Carnegie Mellon University, USA, \\$^3$Mohammed bin Zayed University of AI, UAE}

\begin{document}
\ninept
\maketitle

\begin{abstract}
Automatic Speaker Diarization (ASD) is an enabling technology with numerous applications, which deals with recordings of multiple speakers, raising special concerns in terms of privacy. In fact, in remote settings, where recordings are shared with a server, clients relinquish not only the privacy of their conversation, but also of all the information that can be inferred from their voices.
However, to the best of our knowledge, the development of privacy-preserving ASD systems has been overlooked thus far.
In this work, we tackle this problem using a combination of two cryptographic techniques, Secure Multiparty Computation (SMC) and Secure Modular Hashing, and apply them to the two main steps of a cascaded ASD system: speaker embedding extraction and agglomerative hierarchical clustering. Our system is able to achieve a reasonable trade-off between performance and efficiency, presenting real-time factors of 1.1 and 1.6, for two different SMC security settings.
\end{abstract}


\section{Introduction}
\label{sec:intro}
Automatic Speaker Diarization (ASD) is an enabling technology for many speech-based applications. When combined with Automatic Speech Recognition systems, ASD can provide additional context to the reader, make transcripts clearer or even be used to perform speaker adaptation. On its own, ASD may also allow users to search for and filter segments that correspond to specific speakers, or, in the case of audio diarization, specific audio events. This filtering may be particularly important in multi-speaker audio streams, where the target is a single speaker. In security applications, this speaker may be a potential blacklisted criminal. In clinical interviews, it may be the patient. In language acquisition recordings it may be the child whose linguistic skills are being assessed. The list of potential ASD scenarios is very extensive, ranging from courtrooms, to meetings, sociolinguistic interviews and broadcast news, among others~\cite{tranter2006overview, park2022review}. 

When dealing with large amounts of speech data, when ASD is used as part of a larger system, or even due to the lack of computational resources, it may be useful to delegate this task to an external service.
However, this setting creates a major privacy challenge: the server will have direct access to the user's data.
This means that the voices present in the recording and what is being said, will be available to the server, giving it a very large repository of potentially sensitive information \cite{singh2019profiling}, which the speakers may want to keep private, or may even need to keep private for legal reasons (e.g., to follow privacy regulations such as the EU's GDPR \cite{gdpr}) \cite{nautsch2019preserving}. 
The alternative of having the diarization process run on the user's device is also unattractive, as it would require the service provider to share their model with the user. Considering that ASD models require large amounts of data and high levels of expertise to be developed, sharing them with users would make the service provider potentially lose the value that the model holds. In cascaded ASD models, this is particularly true for the speaker embedding extraction model \cite{teixeira2022towards}.

The above, make this (mostly) unexplored problem -- with the notable exception of \cite{parthasarathi2012wordless} -- particularly interesting. 
In this work we build on our previous contribution on the privacy-preserving extraction of \textit{x-vector} embeddings using Secure Multiparty Computation (SMC) \cite{teixeira2022towards} and extend it to the setting of ASD. 

Specifically, we propose a system that performs the extraction of speaker embeddings and the clustering step in a privacy-preserving way, by leveraging two cryptographic techniques: SMC and Secure Modular Hashing (SMH). The combination of these techniques allows us to protect the service provider's model, particularly the speaker embedding extraction model, while at the same time keeping the speakers' data hidden from the server.

The remainder of this document is organised as follows: in Section \ref{sec:back} we provide the necessary background on SMC and SMH; Section \ref{sec:ppasd} describes the ASD baseline model, and our privacy-preserving system; in Section \ref{sec:exp_setup} we describe the experimental setup; in Section \ref{sec:results} we present and discuss the results obtained; finally, Section \ref{sec:conclusions} presents our conclusions and topics for future work.

\vspace{-0.3cm}
\section{Cryptographic background}
\label{sec:back}

\subsection{Secure Multiparty Computation}
\label{sec:smc}
Secure Multiparty Computation (SMC) is an umbrella term for protocols designed to allow several parties to jointly and securely compute functions over their data, while keeping all inputs private.
SMC protocols are usually built over some form of Secret Sharing \cite{gmw, bgw} or Garbled Circuits \cite{yao, beaver1990round}, and are often combined with cryptographic primitives like Homomorphic Encryption (HE) or Oblivious Transfers (OTs) to perform specific functionalities \cite{lindell2020secure}.
Our privacy-preserving approach will heavily rely on SMC, and particularly on Secret Sharing, which is briefly described below\footnote{For a more in-depth introduction to SMC we direct readers to \cite{lindell2020secure}.}. 

\begin{figure*}[t]
    \centering
\resizebox{0.75\textwidth}{!}{
    \includegraphics{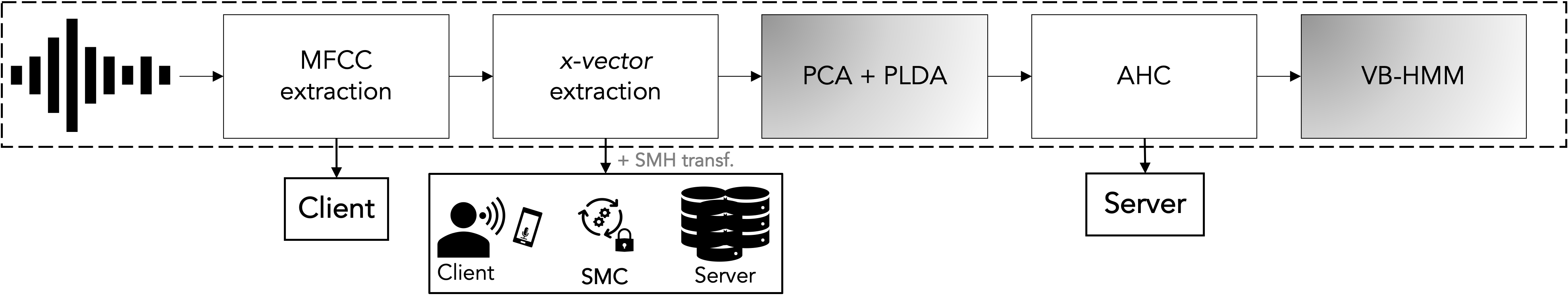}}
    \caption{Block diagram of the original and privacy-preserving systems. Blocks inside the dashed-box correspond to original system; grey-background blocks are skipped in the privacy-preserving system;}
    \label{fig:system}
\end{figure*}
\vspace{-0.3cm}
\subsubsection{Secret Sharing}
\label{sec:smc:secret}
Secret Sharing is a basic primitive for SMC protocols. It allows data owners to represent and share their data with other parties such that each party participating in the computation will only have access to a random-looking \textit{share} (here denoted as $\share{\cdot}$) of the original value. 
Considering an additive \textit{arithmetic} secret sharing scheme in the $n$-party case, a value $x$ shared among several parties by a dealer is defined as:
\begin{equation}
\label{eq:ss}
    x = \share{x}_1 + \share{x}_2 + ... + \share{x}_n,
\end{equation}
\noindent where $\share{x}_1$, ..., $\share{x}_n$ represent random-looking shares of $x$ held by each party, generated as $\share{x}_{n} = x - \sum_{i=1}^{n-1} s_i$, where each $s_i$ is chosen uniformly at random.
When a value is represented in this way, a single party is not able to reconstruct the \textit{secret} without the remaining parties.
This representation allows for parties to interactively compute any operation over their secret data, for both arithmetic and boolean domains. 
For a number of parties strictly larger than two, it is also possible to instantiate more efficient Replicated Secret Sharing (RSS) schemes, wherein instead of holding a single share, each party will hold a set of shares per value \cite{araki2016high}. 

\vspace{-0.2cm}
\subsubsection{Security}
The shared nature of SMC protocols requires making threat assumptions about the participating parties. The threat model of an SMC protocol is extremely important as it significantly affects its security, computational and communication performance.

The most common security (or threat) models include the \textit{semi-honest} and \textit{malicious} models.
The \textit{semi-honest} model is the simplest model possible.
In this model, the adversaries are assumed to follow the established protocol, but are also assumed to pry into and record the data that is visible to them. If all parties follow this behaviour, no party will be able to obtain information other than the one they are allowed to, resulting in very efficient implementations. 
On the other hand, the \textit{malicious} model assumes that adversaries will attempt to thwart the protocol, demanding additional proof that each party is behaving correctly \cite{spdz, spdz2k, mpspdz}. 

In protocols with more than two parties, we can also define security in terms of whether a \textit{majority} of the parties will behave correctly or not -- \textit{honest} vs \textit{dishonest majority}, and whether a subset of parties might \textit{collude} to obtain more information than they are allowed to. 

\vspace{-0.2cm}
\subsection{Secure Modular Hashing}
Secure  Modular  Hashing  (SMH) \cite{SMH}  is  a non-invertible, information-theoretic secure variant of locality-sensitive hashing \cite{LSH}, which projects vectors to a space in such way that, if the Euclidean distance between any pair of vectors in the input space is bellow a certain threshold, their normalised Hamming distance in the new projected space will be proportional to the Euclidean distance in the input space.
Contrarily, if the distance between the two vectors in the input space is larger than a threshold, the distance in the projected space will no longer provide meaningful information, as the Hamming distance will saturate. The relation between the Hamming and Euclidean distances is thus logarithmic shaped -- cf. Fig. 3 in \cite{SMH}.
To accomplish this, in SMH, hash functions are defined as random projections $Q_k$ from $\mathbb{R}^N$ into $(\mathbb{Z}/k)^{M}$, where $\mathbb{Z}/k$ 
is the set of integers from $0$ to $k-1$ and $M$ is the number of hashes, such that $Q_k$ is defined as:
\begin{equation}
\centering
    \begin{split}
    Q_k(x) &= \lfloor A \cdot x + w \rfloor (\mbox{mod k})\mbox{,}
    \end{split}
    \label{eq:smh}
\end{equation}
\noindent with $w \in \mathbb{R^N} \sim \mbox{unif}[0,k]$ and $A \in \mathbb{R}^{N\times M} \sim N\left(0, \frac{1}{\delta^2}I_N\right)$, $M = \mbox{\# \textit{features}} \times \mbox{\# \textit{mpc}}$ (where $mpc$ is the number of \textit{measurements per coefficient}) \cite{sbeCrypto}.
The tuple $(A, w)$ should be treated as the framework's key, and should be kept secret in order to ensure SMH security, because if an adversary does not have access to this key, it cannot meaningfully compare transformed vectors to other vectors they may possess. 

\vspace{-0.2cm}
\section{Privacy-preserving ASD}
\label{sec:ppasd}

\subsection{Baseline system}
\label{sec:baseline}
As a baseline system for our work, we adopted the DIHARD III challenge's baseline \cite{ryant2021dihard} -- Fig. \ref{fig:system}, inside dashed box. Even though better performing systems were later submitted to this challenge (e.g., \cite{wang2021ustc, horiguchi2021hitachi, landini2021but}), its relative simplicity, and the modular nature of this cascaded approach is advantageous to its privacy-preserving implementation as it easily allows the use of different cryptographic methods for different system modules.

The first step in this system is the extraction of Mel Frequency Cepstral Coefficient (MFCC) features from short, overlapping speech frames.
An implicit segmentation is assumed, and \textit{x-vector} embeddings \cite{snyder2018xvectors} are extracted from the resulting speech segments. 
The following step is to perform dimensionality reduction, by applying Principal Component Analysis (PCA) over the zero-centered, whitened, and length normalised speaker embeddings. This is followed by Probabilistic Linear Discriminant Analysis (PLDA) scoring. Agglomerative Hierarchical Clustering (AHC) is then applied to the resulting scores. The baseline system includes a final re-segmentation stage, using a Variational Bayes - Hidden Markov Model (VB-HMM) \cite{diez2018speaker}. 
For simplicity, in this work we only consider oracle Voice Activity Detection (VAD).

\begin{table*}[ht!]
\centering
\caption{Computational and communication costs obtained for the extraction of \textit{x-vectors} and SMH transformation. Values denoted with $^{\$}$ were linearly estimated from a batch size of 700. All results were obtained by averaging over 100 runs.}
\resizebox{0.7\textwidth}{!}{
\begin{tabular}{c|c|c|c|c|c|c}
\toprule
\multirow{2}{*}{Protocol} & \multirow{2}{*}{Security Model} & \multirow{2}{*}{Batch Size} & \multicolumn{2}{c|}{\textit{x-vector} extraction} & \multicolumn{2}{c}{SMH transformation} \\
& & & Time (s) & Comm. (MB) & Time (s) & Comm. (MB) \\\midrule

\multirow{3}{*}{3-party RSS \cite{araki2016high}} & \multirow{3}{*}{HM/SH} 
    & 256  & 73.27 $\pm$ 0.41 & 1,691.03 & 0.90 $\pm$ 0.04 & 12.92 \\
    & & 1024 & 280.05 $\pm$ 2.77$^{\$}$ & 6,564.95$^{\$}$ & 2.88 $\pm$ 0.04 & 72.1 \\ 
    & & 2048 & 560.11 $\pm$ 5.53$^{\$}$ & 13,129.90$^{\$}$ & 4.90 $\pm$ 0.05 & 135.63 \\\midrule

\multirow{3}{*}{4-party RSS \cite{dalskov2021fantastic}} & \multirow{3}{*}{HM/Mal}
        & 256 & 108.50 $\pm$ 1.34 & 5,098.15 & 2.03 $\pm$ 0.04 & 62.89 \\
      & & 1024 & 404.12 $\pm$ 5.25$^{\$}$  & 19,890.61$^{\$}$ & 4.53 $\pm$ 0.04 & 176.84 \\ 
      & & 2048 & 808.25 $\pm$ 10.50$^{\$}$ & 39,781.23$^{\$}$ & 7.20 $\pm$ 0.05 & 331.32 \\ \bottomrule
\end{tabular}
}
\label{tab:results_batch}
\end{table*}
\vspace{-0.2cm}
\subsection{Privacy-preserving system}
Our privacy-preserving system -- composed of the white-background boxes in Fig. \ref{fig:system} -- is a simplified adaptation of the baseline system, consisting of three steps:
\begin{enumerate}
    \item \textbf{Feature extraction} -- to be performed by the client;
    \item \textbf{Speaker embedding extraction} -- to be performed collaboratively by the client and the server using SMC.
    \item \textbf{Agglomerative Hierarchical Clustering} -- to be performed by the server.
\end{enumerate}

The reasoning behind Step 1 was the high computational cost that would be involved in a remote privacy-preserving feature extraction process \cite{thaine2019extracting}. 
It is also important to note that we discard the re-segmentation step, due to its computational complexity, and because it only provides a limited gain to the baseline system \cite{ryant2021dihard}.

Step 2 closely follows the work of \cite{teixeira2022towards}. The use of SMC allows us to protect the speaker embedding model from the client, and the client's data from the server. Similar to \cite{teixeira2022towards}, we assume the existence of trusted parties that will participate in the computation, in order to improve its efficiency. We direct readers to the full paper for a more detailed discussion on the role of external parties in the computation.

Step 3 in our pipeline is the clustering algorithm. We assume that this step should be performed by the server to minimise the computational cost on the client's side, and to allow for a level of flexibility of the overall diarization pipeline (i.e., the server can change the clustering algorithm without having to communicate with the client).

To keep the privacy guarantees provided by Step 2 -- the client does not have access to information about the model, the server does not have access to information about the client's data -- we need to ensure that neither the client or the server have access to the \textit{x-vectors}. However, we also want the server to be able to cluster these vectors.
To solve the latter, we consider the use of SMH. 

Given that SMH is a non-invertible transformation, assuming that the SMH \textit{key} is kept secret by the client and is not re-used, and that the server does not have access to any non-transformed vector in the set, this would allow us to share the set of SMH vectors knowing that: 1) vectors cannot be meaningfully compared to vectors outside the set; 2) the only information the server is able to obtain is the spatial configuration of the set of vectors with regard to each other.

To achieve the above, we also need to ensure that the client does not have access to the \textit{x-vectors} and that the server does not have access to the SMH key.
To this end, we can apply the SMH transformation using SMC, where the key is secret-shared with the server, and eq. \ref{eq:smh} is collaboratively applied to the already secret-shared \textit{x-vectors}.
The client can then send its resulting shares to the server, who can then proceed with the AHC step.
The use of SMH, however, requires that we discard the PCA and PLDA components of the pipeline, as we need to compare vectors directly with the Hamming distance, which will introduce some degradation. Further degradation may also be introduced by cases where the Euclidean distance between vectors is larger than the saturation threshold.

\section{Experimental Setup}
\label{sec:exp_setup}

\subsection{DIHARD III corpus and evaluation}
The DIHARD III challenge dataset is a multi-domain dataset, with development (\textit{dev}) and evaluation (\textit{eval}) partitions, consisting of recordings of 5-10 minute-long samples drawn from 11 domains. 
The \textit{dev} partition includes 254 recordings, while the \textit{eval} partition includes 260 recordings.
In the experiments described in this work, we report our results with regard to the \textit{core} subset of each of these partitions, using the metrics provided for this challenge \cite{ryant2021dihard}.

\vspace{-0.2cm}
\subsection{Speaker embedding extraction}
Our experiments use SpeechBrain's pre-trained \textit{x-vector} model \cite{speechbrain}.
This model follows the architecture of \cite{snyder2018xvectors}. Although models with fewer parameters can be found (e.g. FastResNet-34 \cite{chung2020defence}), the fact that the \textit{x-vector} network is composed of only 7 layers gives it an advantage due to the fatct that, in SMC, non-linear activation layers are much more expensive to compute than linear layers.
The model was trained using the \textit{dev} sets of Voxceleb 1 and 2 \cite{voxceleb2, nagrani2017voxceleb}, achieving 3.2\% Equal Error Rate (EER) on VoxCeleb 1's test set (Cleaned).
\textit{x-vectors} were extracted using 1.5 s windows and 0.25 s shift.

\vspace{-0.3cm}
\subsection{Privacy-preserving implementation}
\label{sec:implementation}
The speaker embedding extraction network and SMH transformation were implemented with the MP-SPDZ library \cite{mpspdz}, using two protocols: the 3-party \textit{semi-honest} RSS scheme of \cite{araki2016high} and the 4-party RSS scheme of \cite{dalskov2021fantastic}, which provides \textit{malicious} (Mal) security in the \textit{honest majority} setting against one corrupted party. These protocols were found to be the most efficient for \textit{x-vector} extraction in \cite{teixeira2022towards}.

For both protocols we used local share conversions and probabilistic truncation to improve efficiency \cite{dalskov2020secure, dalskov2021fantastic}.
Our experiments assume the default parameters for 40-bit security.
We used the library's fixed-point number representation adopting the default configuration of 16 bits for the decimal part, and 15 bits for the fractional part.
All tested protocols perform computations modulo $2^{k}$, where $k = 64$. Experiments were performed on a machine with 24 Intel(R) Xeon(R) CPU E5-2630 v2 @ 2.60GHz processors and 250GB of RAM.
For the SMH transformation we use the following parameters: $k = 2$, $\delta = 15.0$ and $mpc = 4 $.

Considering the fact that we need to extract multiple \textit{x-vectors} from each file to be diarized, we performed experiments using different batch sizes of \textit{x-vectors}: 256, 1024 and 2048. These values were selected based on the statistics of the number of \textit{x-vectors} extracted per file in the DIHARD III \textit{dev} set, encompassing an interval covering roughly 85\% of the data.
However, due to computational limitations, we were unable to directly extract batch sizes larger than 700. As such, the values corresponding to 1024 and 2048 were linearly estimated from the cost of extracting a batch size of 700.
For the SMH transformation, given that the overall computation is lighter, we were able to compute the cost for these batch sizes directly.

\setcounter{table}{2}
\begin{table*}[ht!]
\caption{Results for the Clinical and MapTask domains for the baseline and PRIVADIA systems using task-specific thresholds selected for the \textit{dev} set. Values on the left-hand (resp. right-hand) side of $\rightarrow$ indicate the result obtained for the original (resp. adapted) thresholds.}
\centering
\resizebox{0.65\textwidth}{!}{
\begin{tabular}{l|l|cccc}
\toprule
\multicolumn{1}{c|}{\multirow{2}{*}{\textbf{System}}} & \multicolumn{1}{c|}{\multirow{2}{*}{\textbf{Domain}}} & \multicolumn{2}{c}{Development} & \multicolumn{2}{c}{Evaluation} \\
& & DER (\%) & JER (\%) & DER (\%) & JER (\%) \\ \midrule

\multirow{2}{*}{Simplified baseline} & \multirow{1}{*}{Clinical} & 15.28$\rightarrow$14.7  & 21.71$\rightarrow$21.0 & 15.82$\rightarrow$14.02 & 25.94$\rightarrow$25.27 \\
& \multirow{1}{*}{MapTask}  & 11.01$\rightarrow$10.64 & 18.48$\rightarrow$19.07 & 8.66$\rightarrow$8.53 & 15.55$\rightarrow$15.45 \\ \midrule
\multirow{2}{*}{PPASD} & \multirow{1}{*}{Clinical} & 37.56$\rightarrow$16.65 & 67.36$\rightarrow$25.73 & 39.53$\rightarrow$19.17 & 68.71$\rightarrow$35.46 \\
& \multirow{1}{*}{MapTask} & 24.34$\rightarrow$12.72 & 54.31$\rightarrow$26.99 & 23.61$\rightarrow$15.81 & 59.64$\rightarrow$33.30\\ \bottomrule

\end{tabular}
}
\label{tab:perdomain}
\end{table*}

\vspace{-0.2cm}
\section{Results}
\label{sec:results}

\subsection{Computational and communication costs}
Table \ref{tab:results_batch} presents the computational and communication costs for both the extraction of \textit{x-vectors} and the SMH transformation.
With regard to \textit{x-vector} extraction, we can see that, in the average case, for 1024 \textit{x-vectors}, our method takes $\sim$5 mins and $\sim$6.5 GB for the 3-party setting, and $\sim$7 mins and $\sim$19.5 GB, for the 4-party setting.
Taking into account that a single \textit{x-vector} represents $\sim$0.25 seconds of speech, 1024 \textit{x-vectors} represent roughly 4 mins of speech, without any silence, corresponding to real-time factors of 1.1 and 1.6, for 3- and 4-party RSS, respectively.

When extracting 2048 \textit{x-vectors}, in the 3-party setting, our implementation takes an average of ${\small\sim}$9 minutes, and requires a total of ${\small\sim}$12 GB of communication per party. For the 4-party setting, our method takes ${\small\sim}$13.5 minutes and ${\small\sim}$38 GB of data. In this case, we see that the 4-party protocol starts to become very inefficient.

In terms of the SMH transformation, similar to the \textit{x-vector} extraction, the 3-party RSS setting is more efficient. However, in this case, the added cost of the 4-party setting can be deemed acceptable, particularly if we consider that the overall cost of the SMH transformation is smaller than that of the \textit{x-vector} extraction by close to 2 orders of magnitude, making it negligible in comparison.

\vspace{-0.3cm}
\subsection{Diarization results}
Having discussed how to extract \textit{x-vectors} and apply SMH, our last step is to cluster the transformed vectors, using AHC, and comparing them using the Hamming Distance.
As stated in Section \ref{sec:baseline}, the original baseline system uses PCA reduction, PLDA scoring and re-segmentation, while our system directly clusters the SMH-transformed \textit{x-vectors}. Because of this, and to better assess the degradation introduced by SMH, we provide results with and without these components.
These results can be found in Table \ref{tab:xvectors}, wherein the DIHARD III baseline corresponds to our baseline system, the \textit{Simplified baseline} corresponds to the baseline system without the PCA, PLDA and re-segmentation, and PP-ASD denotes our privacy-preserving system.

From the table, we can see that for the \textit{dev} set, the simplified baseline introduces a degradation of ${\small\sim}$4\% in terms of DER and ${\small\sim}$9\% in terms of JER. Further, by transforming the \textit{x-vectors} with SMH, we add a additional further of ${\small\sim}$2.5\% DER and ${\small\sim}$10\% JER, most likely due to the saturation property of SMH.
Our final system thus introduces a total degradation of ${\small\sim}$7.5\% DER and ${\small\sim}$19\% JER.

\setcounter{table}{1}
\begin{table}[h!]
\centering
\caption{Results obtained for each ASD system.}
\resizebox{\columnwidth}{!}{
\begin{tabular}{lrrrr}
\toprule
\multicolumn{1}{c}{\multirow{2}{*}{System}} & \multicolumn{2}{c}{Development} & \multicolumn{2}{c}{Evaluation}\\
& \multicolumn{1}{c}{DER (\%)} & \multicolumn{1}{c}{JER (\%)} & \multicolumn{1}{c}{DER (\%)} & \multicolumn{1}{c}{JER (\%)} \\ \midrule
DIHARD III baseline \cite{ryant2021dihard} & 20.25 & 46.02 & 20.65 & 47.74\\
Simplified baseline & 25.36 & 55.75 & 25.15 & 54.57 \\
PP-ASD              & 27.95 & 65.52 & 29.58 & 67.72 \\
\bottomrule
\end{tabular}
}
\label{tab:xvectors}
\end{table}

When looking at the \textit{eval} results, while the degradation remains similar from the original to the simplified baseline, we observe a stronger degradation -- ${\small \sim}9\%$ DER and ${\small \sim}20\%$ JER -- when applying SMH. We hypothesise that this is due to the non-linear relation between the Euclidean and Hamming distances, making our system more sensitive to the clustering threshold optimised for the \textit{dev} set.

\subsection{Per-domain analysis}
To have a better understanding of the effects of SMH with regard to the simplified baseline system's performance, we decided to analyse the system's results in a per-domain basis. 
We found that, even though the overall degradation is close to 2.5\% in terms of DER and 10\% in terms of JER for most domains, for some domains, namely for the Clinical and MapTask domains, the degradation introduced by SMH was disproportionate with regard to the baseline: an absolute degradation of 22\% (24\%) DER and 46\% (43\%) JER for the Clinical domain, and of 13\% (15\%) DER and 36\% (44\%) JER for the MapTask domain, in the \textit{dev} (\textit{eval}) set. 
We again claim that this is a direct effect of the sensitivity to the threshold selection, due to the saturation effect introduced by SMH, and the fact that a single threshold was selected for all domains. 

To verify this hypothesis, we decided to adjust the threshold for each of the \textit{dev} set domains.
We did this for both the simplified baseline and our final private system, to provide a fair comparison.
The results for this experiment are provided in Table \ref{tab:perdomain}, for the two above mentioned domains.
We can see that by adjusting the threshold, our simplified baseline results show only small levels of improvement. Contrarily, for the privacy-preserving system, the results highly improve: $\sim$21\% (20\%) DER and $\sim$42\% (33\%) JER for the Clinical domain and $\sim$12\% (8\%) DER and $\sim$28\% (26\%) JER for the MapTask domain, for the \textit{dev} (\textit{eval}) set.
When comparing these results to the new baseline results, we see that the SMH-introduced degradation is close to 2.5\% DER and less than 10\% JER in both settings, which is in line with the average degradation, thus proving our hypothesis.

The above shows that the privacy-preserving system is more sensitive to domain changes, making it less reliable than the baseline system. However, in a real world application, it is reasonable to assume that one can ask a potential user to define the setting in which the recording under evaluation was made, so that the system can use a domain specific threshold to optimise performance.
\vspace{-0.3cm}
\section{Conclusions}
\label{sec:conclusions}

In this work we have presented the first implementation of a privacy-preserving speaker diarization system using existing cryptographic techniques.
The contributions of this work are not limited to the setting of ASD, as we introduce a approach to apply SMH in a privacy-preserving way, using SMC, which has potential applications in the area of template protection \cite{SMH}.

This system still has limitations, both in terms of ASD performance and computational cost, but we foresee many possible improvements.
Considering that the computational bottleneck is the \textit{x-vector} extraction, reducing the size of this model could help improve efficiency, while exploring other techniques for the SMH-based clustering could help mitigate the degradation of the results.

Future work could also deal with privacy-preserving feature extraction, VAD, re-segmentation or overlap detection algorithms, all of which are applicable to speech tasks beyond ASD.

\bibliographystyle{IEEEbib}
\bibliography{mybib}

\end{document}